\documentclass{PoS}
\usepackage{amsmath}
\usepackage{subfigure}

\newcommand{\be}{\begin{equation}}
\newcommand{\ee}{\end{equation}}
\newcommand{\bea}{\begin{eqnarray}}
\newcommand{\eea}{\end{eqnarray}}

\title{$|V_{us}|$ from $K$ decays in theory}

\ShortTitle{$|V_{us}|$ from $K$ decays in theory}

\author{Silvano Simula

\\

\it Istituto Nazionale di Fisica Nucleare, Sezione di Roma Tre, Rome, Italy.\\ Email: \email{simula@roma3.infn.it}

}

\abstract{Leptonic and semileptonic kaon decays represent till now the golden modes for the extraction of the Cabibbo angle from experiments, provided the relevant hadronic quantities, namely the ratio of the kaon and pion leptonic decay constants, $f_K / f_\pi$, and the semileptonic vector form factor at zero four-momentum transfer, $f_+(0)$, are computed accurately from QCD. In the last years, using large-scale lattice QCD simulations, the determination of both $f_K / f_\pi$ and $f_+(0)$ has reached an impressive level of precision ($\approx 0.3 \%$), which is expected to be further improved in the near future. However, at a permille level of precision both electromagnetic and strong isospin-breaking effects cannot be neglected anymore. In this contribution a new lattice strategy aiming at determining QCD+QED effects in the inclusive leptonic decay rates of charged pseudoscalar mesons is briefly illustrated, and the preliminary results obtained in the case of $\pi_{\ell 2}$ and $K_{\ell 2}$ decays are presented. As for the semileptonic $K_{\ell 3}$ decays, it is pointed out that the determination of the relevant form factors in the full kinematical range covered by the experiments is necessary for a more stringent test of the Standard Model. In this respect the perspectives based on the only existing results from the European Twisted Mass Collaboration (ETMC) are quite encouraging.}

\FullConference{9th International Workshop on the CKM Unitarity Triangle\\
		28  November - 3 December 2016\\
		Tata Institute for Fundamental Research (TIFR), Mumbai, India}

\begin{document}

\section{Introduction}
\label{sec:intro}

Since several decades the leptonic and semileptonic decays of kaons represent the golden modes for the determination of the Cabibbo-Kobayashi-Maskawa (CKM) matrix element $V_{us}$.
According to the $V - A$ structure of the weak current in the Standard Model (SM) the decay rate of the above processes should provide the same result for $|V_{us}|$, once the relevant hadronic quantities, namely the ratio of the kaon and pion leptonic decay constants, $f_K / f_\pi$, and the semileptonic vector form factor at zero four-momentum transfer, $f_+(0)$, are determined starting from our fundamental theory of the strong interactions, i.e.~from QCD.

As it is well known, such a task can be properly carried out by large-scale lattice QCD simulations.
The history of the efforts in predicting both $f_K / f_\pi$ and $f_+(0)$ is nicely summarised in the recent reviews \cite{FLAG_old,FLAG} of the Flavor Lattice Averaging Group (FLAG), showing that the determination of $f_K / f_\pi$ and $f_+(0)$ has reached an impressive degree of precision ($\approx 0.3 \%$ \cite{FLAG}), such that both electromagnetic (e.m.) and strong isospin-breaking (IB) effects cannot be neglected anymore.

In the past few years accurate lattice results including e.m.~effects have been obtained for the hadron spectrum, as in the case of the neutral-charged mass splittings of light pseudoscalar (PS) mesons and baryons (see, e.g., Refs.~\cite{deDivitiis:2013xla,Borsanyi:2014jba}).
However, while in the calculation of e.m.~effects in the hadron spectrum no infrared (IR) divergencies can appear, in the case of other hadronic quantities, like the decay amplitudes, IR divergencies are present and can be cancelled out in the physical observable only by summing up diagrams containing both real and virtual photons \cite{BN37}.
This is the case of the leptonic $\pi_{\ell 2}$ and $K_{\ell 2}$ and of the semileptonic $K_{\ell 3}$ decay rates, which play a crucial role for an accurate determination of the CKM entries $|V_{us} / V_{ud}|$ and $|V_{us}|$. 

In this contribution we discuss two novelties:
\begin{itemize}
\item a new strategy has been proposed recently \cite{Carrasco:2015xwa} in order to determine on the lattice the inclusive decay rate of a charged PS-meson into either a final $\ell^\pm \bar{\nu}_\ell$ pair or a final $\ell^\pm \bar{\nu}_\ell \gamma$ state. 
In Section \ref{sec:Kl2} the new approach will be briefly presented together with the preliminary results obtained in the case of $\pi_{\ell 2}$ and $K_{\ell 2}$ decays \cite{Lubicz:2016mpj}. 
\item the momentum dependence of the semileptonic form factors in the full kinematical range covered in the $K_{\ell 3}$ experiments has been determined only recently by ETMC \cite{Carrasco:2016kpy}. 
The results obtained at the physical pion point will be presented in Section \ref{sec:Kl3}, showing that they are very encouraging for obtaining in the next future a more stringent test of the SM from $K_{\ell 3}$ decays.
\end{itemize}

\section{$|V_{us} / V_{ud}|$ from $\pi_{\ell 2}$ and $K_{\ell 2}$ decays}
\label{sec:Kl2}

In the PDG review \cite{PDG} the inclusive decay rate $\Gamma(PS^+ \to \ell^+ \nu_\ell [\gamma])$ is written as
 \be
      \Gamma(PS^+ \to \ell^+ \nu_\ell [\gamma]) =  \frac{G_F^2}{8 \pi} |V_{q_1 q_2}|^2 m_\ell^2 \left( 1 - \frac{m_\ell^2}
            {M_{PS}^2} \right)^2 f_{PS^+}^2 M_{PS} (1 + \delta_{e.w.} + \delta_{e.m.}^{PS^+}) ~ ,
     \label{eq:Gamma_PDG}
 \ee
where $\delta_{e.w.}$ is a universal short-distance electroweak correction ($\delta_{e.w.} \simeq 0.0232$), $\delta_{e.m.}^{PS^+}$ represents the e.m.~corrections, estimated through the Chiral Perturbation Theory (ChPT) with low-energy constants (LECs) parameterizing structure-dependent hadronic contributions, $ f_{PS^+}$ is the decay constant of the charged PS-meson including the IB effects generated in QCD by the $u$- and $d$-quark mass difference, and $M_{PS}$ is the charged PS-meson mass including both e.m.~and strong IB corrections.
Adopting LECs motivated by large-$N_c$ methods one has $\delta_{e.m.}^{K^+} - \delta_{e.m.}^{\pi^+} = -0.0069 (17)$ \cite{Cirigliano:2011tm}, so that the experimental value of the ratio of the $K_{\ell 2}$ and $\pi_{\ell 2}$ decay rates translates into \cite{FLAG,PDG}
 \be
     \frac{\Gamma(K^+ \to \ell^+ \nu_\ell [\gamma])}{\Gamma(\pi^+ \to \ell^+ \nu_\ell [\gamma])} \qquad \Longrightarrow \qquad 
         \frac{|V_{us}|}{|V_{ud}|} \frac{f_{K^+}}{f_{\pi^+}} = 0.2760 (4) ~ .
 \ee
According to the last FLAG review \cite{FLAG} the lattice average for $f_{K^+} / f_{\pi^+}$ is equal to $f_{K^+} / f_{\pi^+} = 1.193 (3)$ with $N_f = 2+1+1$ dynamical quarks. 
The precision is at the level of $\simeq 0.3 \%$, which is already comparable with the uncertainty of the model-dependent ChPT prediction for $\delta_{e.m.}^{K^+} - \delta_{e.m.}^{\pi^+}$ ($\simeq 0.2 \%$).
Thus, a lattice evaluation of the leptonic decay rates including both QCD and QED is mandatory and in this respect a new strategy has been proposed in Ref.~\cite{Carrasco:2015xwa}.
There the inclusive rate $\Gamma(PS^+ \to \ell^+ \nu_\ell [\gamma])$ is expressed as
 \be
     \Gamma(PS^\pm \to \ell^\pm \nu_\ell [\gamma]) = \mbox{lim}_{L \to \infty} [\Gamma_0(L) -  \Gamma_0^{pt}(L)] + 
     \mbox{lim}_{\mu_\gamma \to 0} [\Gamma_0^{pt}(\mu_\gamma) + \Gamma_1^{pt}(\Delta E_\gamma, \mu_\gamma)] ~ ,
     \label{eq:Gamma}
 \ee
where the subscripts $0$ and $1$ indicate the number of photons in the final state, while the superscript ``pt'' denotes the point-like treatment of the decaying PS-meson.
In the r.h.s.~of Eq.~(\ref{eq:Gamma}) the terms $\Gamma_0(L)$ and $\Gamma_0^{pt}(L)$ can be evaluated on the lattice using the lattice size $L$ as an IR regulator.
Their difference is free from IR divergencies and therefore the limit $L \to \infty$ can be performed obtaining a result independent on the specific IR regularization \cite{Carrasco:2015xwa}.
In a similar way the terms $\Gamma_0^{pt}(\mu_\gamma)$ and $\Gamma_1^{pt}(\Delta E\gamma, \mu_\gamma)$ can be calculated perturbatively using a photon mass $\mu_\gamma$ as an IR regulator.
Their sum is free from IR divergencies thanks to the Bloch-Nordsieck mechanism \cite{BN37}, so that the limit $\mu_\gamma \to 0$ can be performed obtaining again a result independent on the specific IR regularization.

Thus, the correction to the tree-level decay rate is given by
 \be
    \delta R_{PS^+} \equiv  \delta_{e.w.} + \delta_{e.m.}^{PS^+} + \delta_{SU(2)}^{PS^+} = \alpha_{em} \frac{2}{\pi} 
         \mbox{log}\left( \frac{M_Z}{M_W} \right) + 2 \delta \left[ \frac{A_{PS}(L) - A_{PS}^{pt}(L)}{A_{PS}^{(0)}} \right] + 
         \delta \Gamma^{(pt)}(\Delta E_\gamma) ~ ,
    \label{eq:RPS}
 \ee
where $\delta \Gamma^{(pt)}(\Delta E_\gamma)$ can be read off from Eq.~(51) of Ref.~\cite{Carrasco:2015xwa} and $A_{PS}^{(0)} \equiv \langle 0 | \bar{q}_2 \gamma_0 \gamma_5 q_1 | PS \rangle = f_{PS}^{(0)} M_{PS}$ is the QCD axial amplitude, with $f_{PS}^{(0)}$ being the PS meson decay constant known in pure QCD (i.e., without e.m.~and strong IB corrections).

\begin{figure}[htb!]
\parbox{10.0cm}{~~ \includegraphics[scale=0.55]{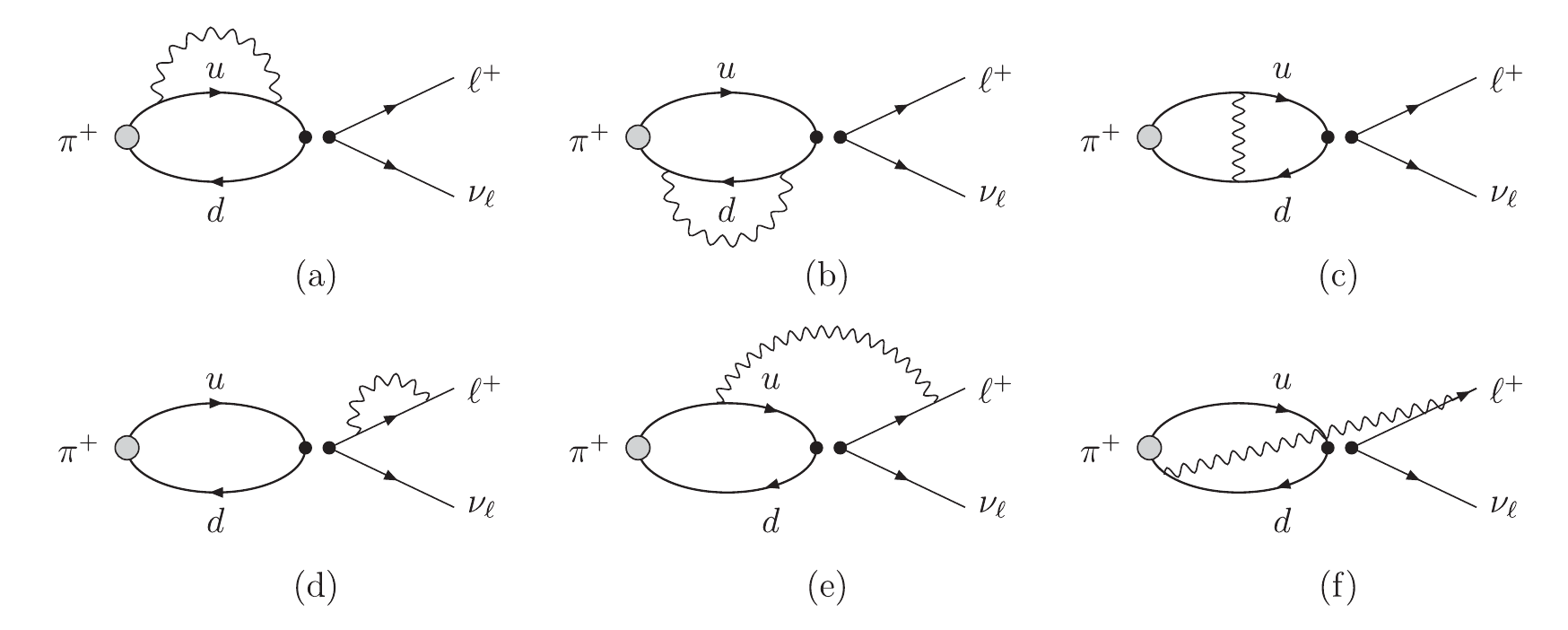}}$~$\parbox{5.0cm}{
\caption{\it \footnotesize Connected diagrams contributing at $O(\alpha_{em})$ to the axial amplitude $\delta A_{\pi^+}$ for the decay $\pi^+ \to \ell^+ \nu$.}
\label{fig:diagrams}
}
\end{figure}

Adopting the quenched QED approximation, which neglects the sea-quark electric charges, the quantity $\delta A_{PS}(L)$ has been evaluated in Ref.~\cite{Lubicz:2016mpj}, adopting the ETMC gauge ensembles with $N_f = 2+1+1$ dynamical quarks and considering the connected diagrams shown in Fig.~\ref{fig:diagrams} and the one corresponding to the insertion of the isovector scalar density \cite{deDivitiis:2013xla}.

\begin{figure}[htb!]
\parbox{7.0cm}{\includegraphics[scale=0.35]{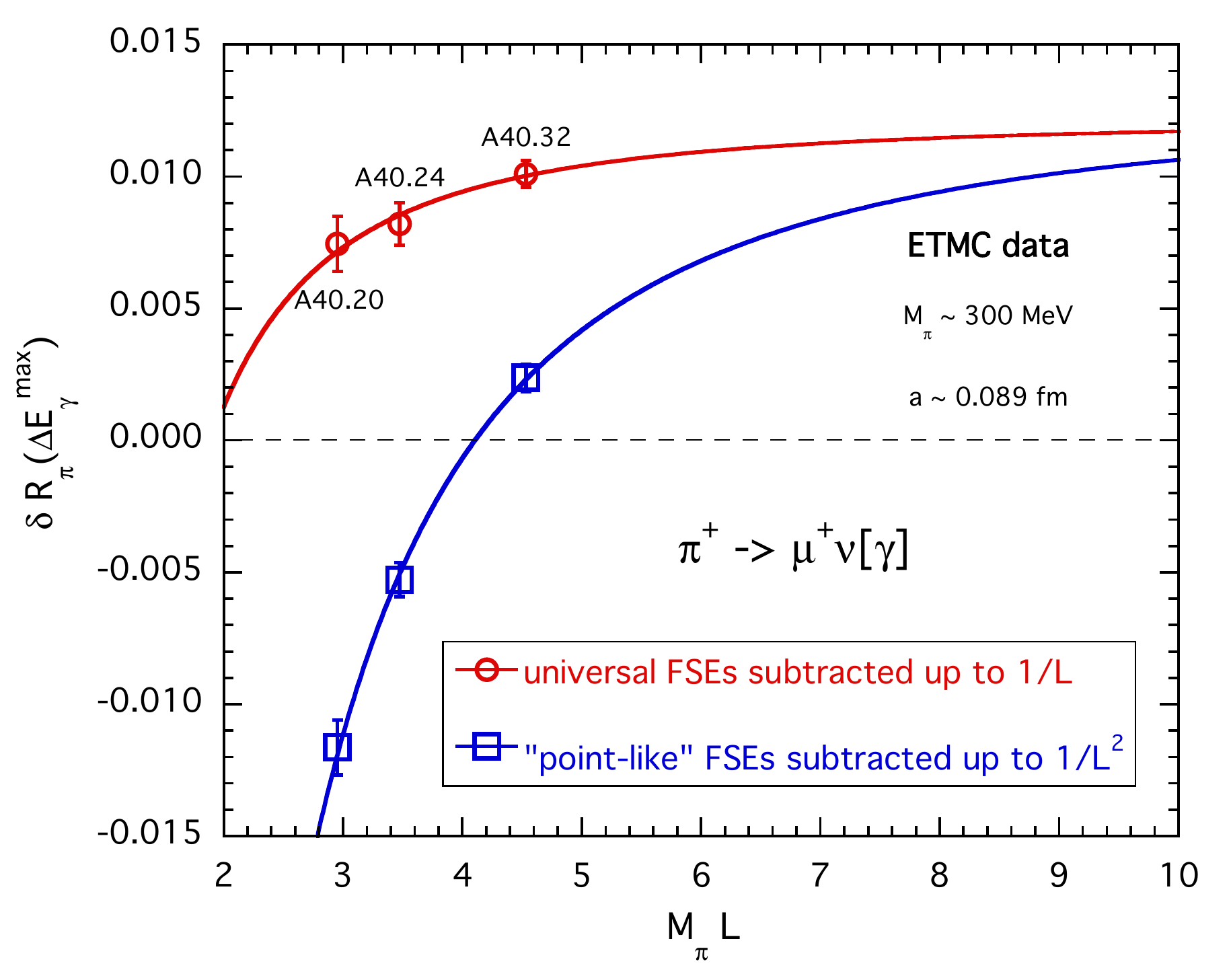}}$~$\parbox{8.0cm}{
\caption{\it \footnotesize Results for the correction $\delta R_\pi(\Delta E_\gamma^{max})$ to the decay $\pi^+ \to \mu^+ \nu [\gamma]$ (see Eq.~(\protect\ref{eq:RPS})) for the ETMC gauge ensembles A40.20, A40.24 and A40.32 corresponding to the same lattice spacing ($a \simeq 0.089$ fm) and pion mass ($M_\pi \sim 300$ MeV), but different lattice sizes (see Ref.~\cite{Carrasco:2014cwa}). The red points correspond to the subtraction of the universal FSEs, i.e.~up to order $O(1/L)$ in Eq.~(\protect\ref{eq:APS_pt}), while the blue squares include also the subtraction of the ``point-like'' term $b_2 / (M_\pi L)^2$. The solid lines are the results of the simple fit $a + b / L^2$ with $a$ and $b$ being free parameters. Note that the two fits agree with each other for $L \to \infty$. \vspace{0.5cm}}
\label{fig:FSE}
}
\end{figure}

In Eq.~(\ref{eq:RPS}) the term $A_{PS}^{pt}(L)$ corresponds to the virtual photon emissions from a point-like meson using the lattice size $L$ as the IR regulator.
Such a quantity has been calculated in Ref.~\cite{Tantalo:2016vxk}, obtaining the result
 \be
     \delta \left[ \frac{A_{PS}^{pt}(L)}{A_{PS}^{(0)}} \right] = b_{IR} ~ \mbox{log}(M_{PS} L) + b_0 + \frac{b_1}{M_{PS} L} + 
                                                                                           \frac{b_2}{M_{PS}^2 L^2} + \frac{b_3}{M_{PS}^3 L^3} + 
                                                                                           O(e^{-M_{PS} L}) ~ ,
       \label{eq:APS_pt}
 \ee
where the coefficients $b_j$ ($j = IR, 0, 1, 2, 3$) depend on the mass ratio $m_\ell / M_{PS}$.
The relevant point is that structure-dependent finite size effects (FSEs) start only at order $O(1/L^2)$, i.e.~all the terms up to $O(1/L)$ in Eq.~(\ref{eq:APS_pt}) are ``universal'' \cite{Tantalo:2016vxk}.
The FSE subtraction (\ref{eq:APS_pt}) is illustrated in Fig.~\ref{fig:FSE} in the case of the decay $\pi^+ \to \mu^+ \nu [\gamma]$ for $\Delta E_\gamma = \Delta E_\gamma^{max} = M_\pi (1 - m_\mu^2 / M_\pi^2) / 2 \simeq 30$ MeV. 
It can be seen that residual FSEs are still visible in the subtracted lattice data.

\begin{figure}[htb!]
\begin{center}
\includegraphics[scale=0.80]{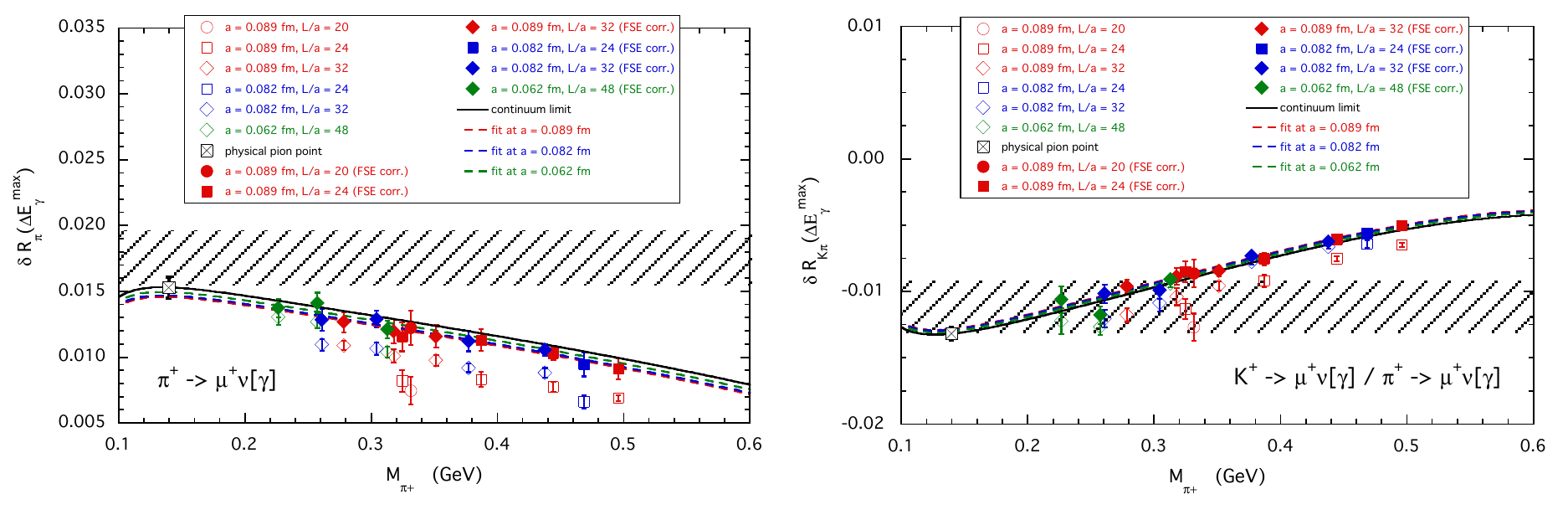}
\end{center}
\vspace{-0.75cm}
\caption{\it \footnotesize Results for the corrections $\delta R_\pi$ (left panel) and $\delta R_{K \pi} \equiv \delta R_K - \delta R_\pi$ (right panel) obtained after the subtraction of the ``universal'' FSE terms of Eq.~(\protect\ref{eq:APS_pt}) (open markers). The full markers correspond to the lattice data of Ref.~\cite{Lubicz:2016mpj} corrected also for the residual FSEs obtained in the fitting procedure. The dashed lines represent the results in the infinite volume limit at each value of the lattice spacing, while the solid lines are the results in the continuum limit. The crosses represent the values $\delta R_\pi^{phys}$ and $\delta R_{K \pi}^{phys}$ at the physical point. The shaded areas correspond respectively to the values $0.0176(21)$ and $-0.0112(21)$ at 1-sigma level, obtained using ChPT \protect\cite{Cirigliano:2011tm,Rosner:2015wva}.}
\label{fig:PSplus}
\end{figure}

The results obtained in Ref.~\cite{Lubicz:2016mpj} for the corrections $\delta R_\pi$ and $\delta R_{K \pi} \equiv \delta R_K - \delta R_\pi$ are shown in Fig.~\ref{fig:PSplus}, where all photon energies are included (i.e.~$\Delta E_\gamma = \Delta E_\gamma^{max} = M_{PS} (1 - m_\mu^2 / M_{PS}^2) / 2$), since the experimental data on $\pi_{\ell 2}$ and $K_{\ell 2}$ decays are fully inclusive. 
The combined chiral, continuum and infinite volume extrapolations are performed using different fitting functions and FSE subtractions in order to estimate the systematic errors. 
After averaging all the results one finally gets at the physical pion point 
 \bea
      \label{eq:Rpi_phys}
      \delta R_\pi^{phys} & = & + 0.0169 ~ (8)_{stat + fit} ~ (11)_{chiral} ~ (7)_{FSE} ~ (2)_{a^2} = + 0.0169 ~ (15) ~ , \\
      \label{eq:RKpi_phys}
      \delta R_{K \pi}^{phys} & = & - 0.0137 ~ (11)_{stat + fit} ~ (6)_{chiral} ~ (1)_{FSE} ~ (1)_{a^2} = - 0.0137 ~ (13) ~ ,
 \eea
where the errors do not include the QED quenching effects.
The results (\ref{eq:Rpi_phys})-(\ref{eq:RKpi_phys}) can be compared with the corresponding ChPT predictions $0.0176(21)$ and $-0.0112(21)$ \cite{Cirigliano:2011tm,Rosner:2015wva}.

A crucial condition for the strategy of Ref.~\cite{Carrasco:2015xwa} is that the maximum energy of the emitted photon, $\Delta E_\gamma$, has to be small enough in order not to resolve the internal structure of the decaying PS-meson.
Note that the corresponding form factors would describe a process which is not related directly to the one the CKM entry can be extracted from.
Thus, the experimental determination of the photon spectrum in $K_{\ell 2}$ decays is required.

\section{$|V_{us}|$ from $K_{\ell 3}$ decays}
\label{sec:Kl3}

The semileptonic decay rate $\Gamma(K^{+,0} \to \pi^{0,-} \ell^+ \nu_\ell [\gamma])$ can be written as \cite{PDG}
 \be
      \Gamma(K^{+,0} \to \pi^{0,-} \ell^+ \nu_\ell [\gamma]) =  \frac{G_F^2 M_{K^{+,0}}^5}{192 \pi^3} C_{K^{+,0}}^2 
          |V_{us} f_+^{K^0 \pi^-}(0)|^2 I_{K \ell} ~ (1 + \delta_{e.w.} + \delta_{e.m.}^{K^{+,0} \ell} + \delta_{QCD}^{K^{+,0} \pi}) ~ ,
     \label{eq:Gamma_Kl3_PDG}
 \ee
where $C_{K^{+,0}}$ is a Clebsh-Gordan coefficient, $I_{K \ell}$ is the phase-space integral sensitive to the momentum dependence of the semileptonic vector and scalar form factors, $f_+^{K^0 \pi^-}(0)$ is the vector form factor at zero four-momentum transfer, $\delta_{e.m.}^{K^{+,0} \ell}$ and $\delta_{QCD}^{K^{+,0} \pi}$ represent the e.m.~and strong IB corrections, respectively.
The latter have been estimated for the various charged and neutral kaon decay modes by means of ChPT \cite{Cirigliano:2008wn}.
The nice consistency among the various decay modes allow to translate the experimental value of the inclusive decay rates into \cite{FlaviaNet,Moulson:2014cra}
 \be
     \Gamma(K^{+,0} \to \pi^{0,-} \ell^+ \nu_\ell [\gamma]) \qquad \Longrightarrow \qquad |V_{us}| f_+^{K^0 \pi^-}(0) = 0.2165 (4) ~ .
 \ee
In the case of $N_f = 2+1+1$ dynamical quarks the lattice average for $f_+^{K^0 \pi^-}(0)$ is equal to $f_+^{K^0 \pi^-}(0) = 0.9706 (27)$ \cite{FLAG}. 
The precision is at the level of $\simeq 0.3 \%$, which is already comparable with the uncertainties of the ChPT predictions  \cite{Cirigliano:2008wn,FlaviaNet} for $\delta_{e.m.}^{K^{+,0} \ell}$ and $\delta_{QCD}^{K^{+,0} \pi}$ ($\simeq 0.1 \%$ and $\simeq 0.4 \%$, respectively).
Thus, a lattice evaluation of the $K_{\ell 3}$ decay rate including both QCD and QED effects is called for.
While the application of the strategy of Ref.~\cite{Carrasco:2015xwa} to $K_{\ell 3}$ decays is still in progress, we want to point out the importance of the lattice evaluation of the phase-space integral $I_{K \ell}$, which requires the determination of the semileptonic vector and scalar form factors, $f_+(q^2)$ and $f_0(q^2)$, in the full kinematical range covered by experiments.
Till now the momentum dependence of the $K_{\ell 3}$ form factors has been determined only by ETMC \cite{Carrasco:2016kpy}.

\begin{figure}[htb!]
\begin{center}
\includegraphics[scale=0.27]{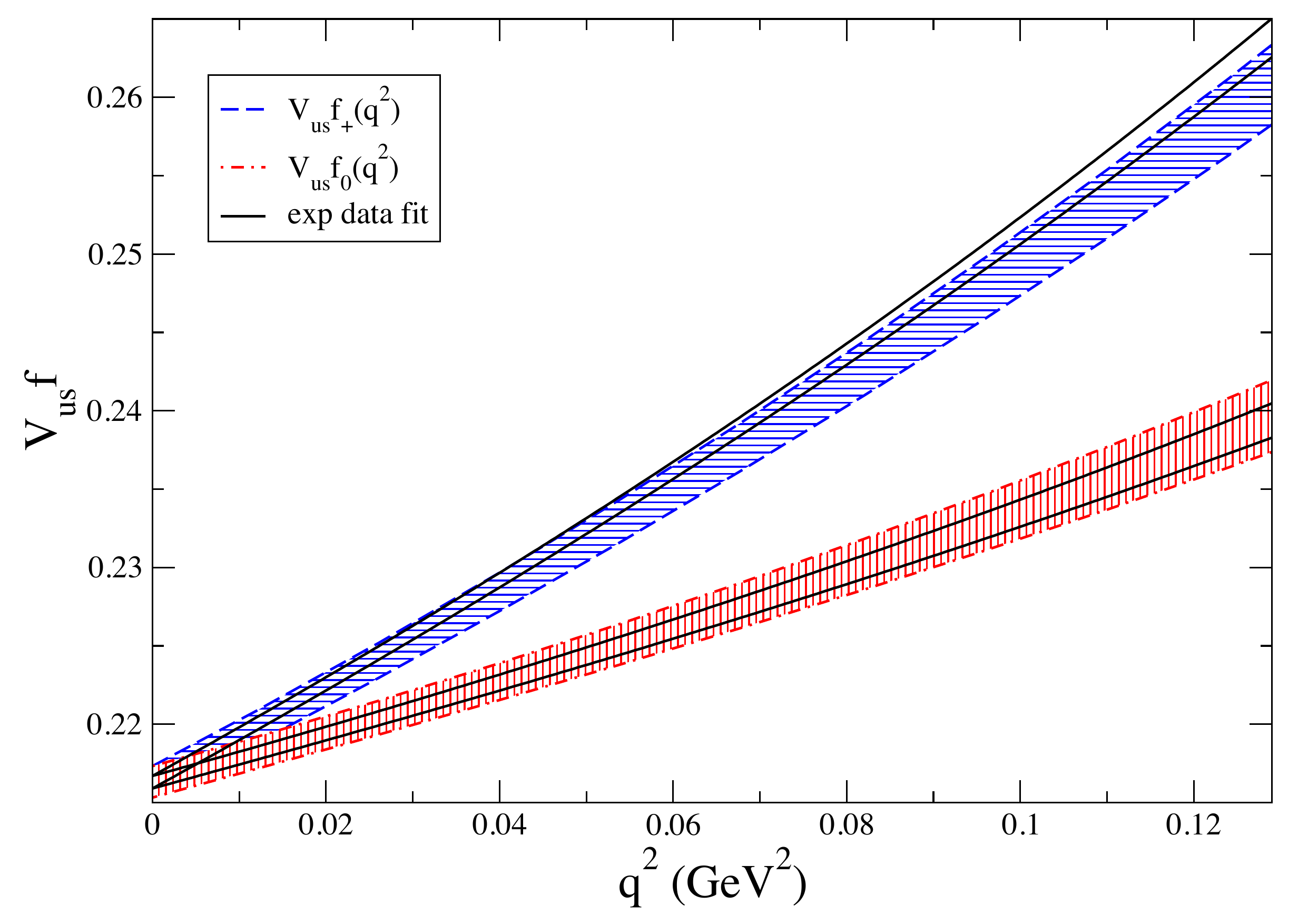} ~
\includegraphics[scale=0.38]{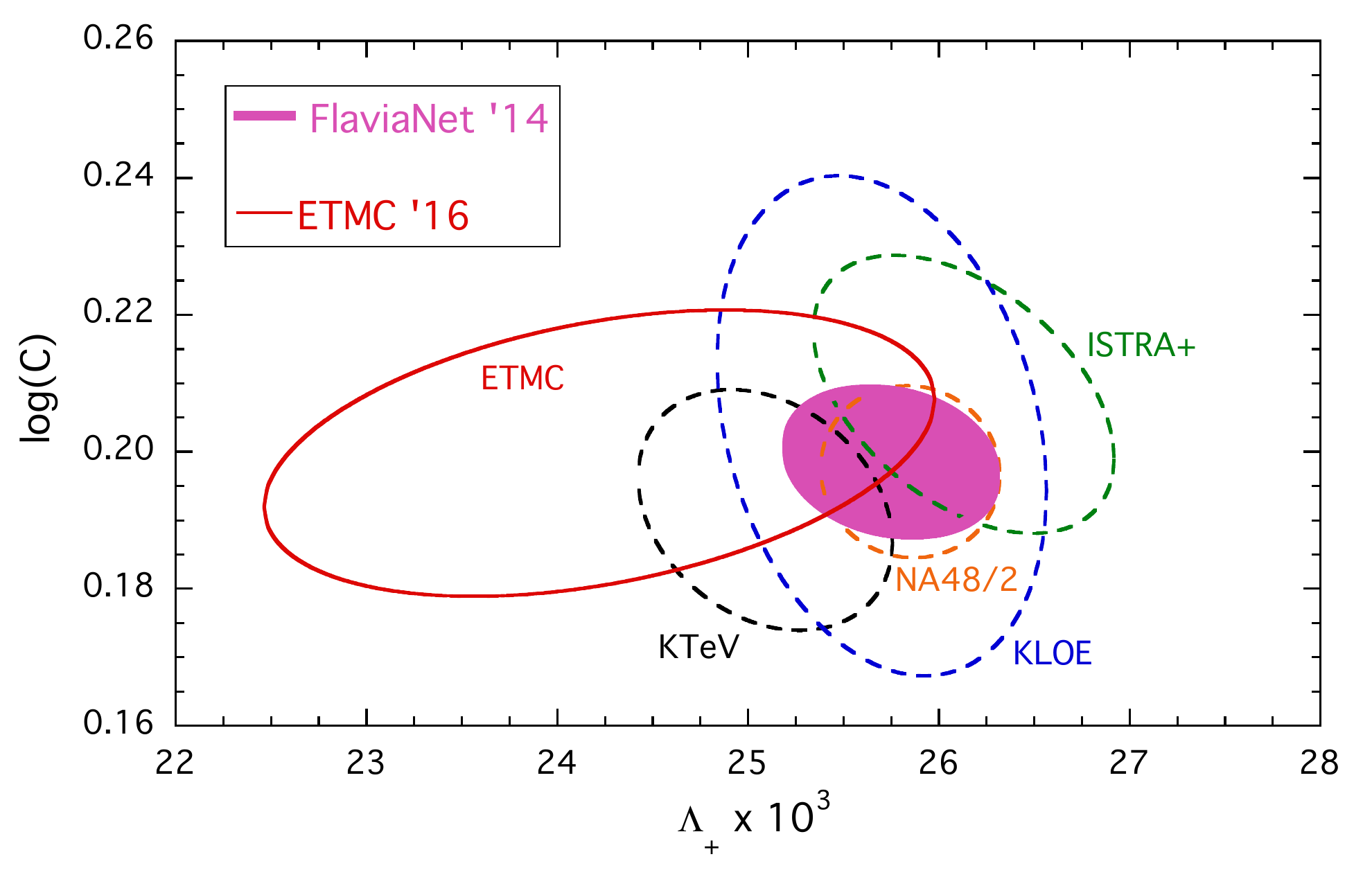}
\end{center}
\vspace{-0.75cm}
\caption{\it \footnotesize Left panel: results for the vector (blue area) and scalar (red area) form factors, obtained in Ref.~\cite{Carrasco:2016kpy} at the physical pion point including both statistical and systematic uncertainties, and multiplied by $|V_{us}| = 0.2230$, versus $q^2$ in the range between $q^2 = 0$ and the physical kinematical end-point $q^2 = q_{max}^2 \simeq 0.129$ GeV$^2$. The black solid lines represent the results of the dispersive fit of the experimental data performed in Ref.~\cite{Moulson:2014cra}. Right panel: comparison of the information for the dispersive parameters $\Lambda_+$ and $\rm{log}(C)$ obtained in Ref.~\cite{Carrasco:2016kpy} (solid ellipse) with the corresponding results of the $K_{\ell 3}$ experiments KTeV, KLOE, NA48/2 and ISTRA+ (dashed ellipses), taken from Refs.~\cite{FlaviaNet,Moulson:2014cra}, and with the updated FlaviaNet average \cite{Moulson:2014cra} (full ellipse). All the ellipses represent contours corresponding to a $68\%$ likelihood.}
\label{fig:Kl3}
\end{figure}

The results obtained at the physical pion point in Ref.~\cite{Carrasco:2016kpy} are shown in Fig.~\ref{fig:Kl3} and compared with the outcome of the analysis of the experimental data \cite{Moulson:2014cra} based on the dispersive representation of Ref.~\cite{Bernard:2009zm}.
The latter depends on two parameters, $\Lambda_+$ and $C$, which represent respectively the slope of the vector form factor $f_+(q^2)$ at $q^2 = 0$ (in units of $M_\pi^2$) and the scalar form factor $f_0(q^2)$ at the (unphysical) Callan-Treiman point \cite{Callan:1966hu} $q^2 = q_{CT}^2 \equiv M_K^2 - M_\pi^2$ divided by $f_+(0)$.
It can be seen that the precision of the lattice results is not far from those of each separate experiments.
Thanks to improvements foreseeable in the next future, the perspectives are quite encouraging for obtaining from lattice QCD a more stringent test of the SM in $K_{\ell 3}$ decays.


\begin{thebibliography}{99}

\bibitem{FLAG_old}
G.~Colangelo {\it et al.},
  Eur.\ Phys.\ J.\ C {\bf 71} (2011) 1695
  [arXiv:1011.4408 [hep-lat]].
  S.~Aoki {\it et al.},
  Eur.\ Phys.\ J.\ C {\bf 74} (2014) 9,  2890
  [arXiv:1310.8555 [hep-lat]].

\bibitem{FLAG}
  S.~Aoki {\it et al.},
  arXiv:1607.00299 [hep-lat]
   and the update at http://itpwiki.unibe.ch/flag/index.php.
\bibitem{deDivitiis:2013xla}
  G.~M.~de Divitiis {\it et al.},
  Phys.\ Rev.\ D {\bf 87} (2013) no.11,  114505
  [arXiv:1303.4896 [hep-lat]];
  JHEP {\bf 1204} (2012) 124
  [arXiv:1110.6294 [hep-lat]].
 
 \bibitem{Borsanyi:2014jba}
  S.~Borsanyi {\it et al.},
  Science {\bf 347} (2015) 1452
  [arXiv:1406.4088 [hep-lat]].
  
\bibitem{BN37}
 F.~Bloch and A.~Nordsieck, Phys.\ Rev.\ {\bf 52} (1937) 54.

\bibitem{Carrasco:2015xwa}
  N.~Carrasco {\it et al.},
  Phys.\ Rev.\ D {\bf 91} (2015) no.7,  074506
  [arXiv:1502.00257 [hep-lat]].

\bibitem{Lubicz:2016mpj}
  V.~Lubicz {\it et al.},
  arXiv:1610.09668 [hep-lat].

\bibitem{Carrasco:2016kpy}
  N.~Carrasco {\it et al.} [ETM Coll.],
  Phys.\ Rev.\ D {\bf 93} (2016) no.11, 114512
  [arXiv:1602.04113 [hep-lat]].

\bibitem{PDG}
 C.~Patrignani {\it et al.} [Particle Data Group],
  Chin.\ Phys.\ C {\bf 40} (2016) no.10, 100001.

\bibitem{Cirigliano:2011tm}
  V.~Cirigliano and H.~Neufeld,
  Phys.\ Lett.\ B {\bf 700} (2011) 7
  [arXiv:1102.0563 [hep-ph]].

\bibitem{Tantalo:2016vxk}
  N.~Tantalo {\it et al.}, 
  arXiv:1612.00199 [hep-lat].
  See also: V.~Lubicz {\it et al.},
  arXiv:1611.08497 [hep-lat].

\bibitem{Carrasco:2014cwa}
  N.~Carrasco {\it et al.} [ETM Coll.],
  Nucl.\ Phys.\ B {\bf 887} (2014) 19
  [arXiv:1403.4504 [hep-lat]].

\bibitem{Rosner:2015wva}
  J.~L.~Rosner {\it et al.},
  Submitted to: Particle Data Book
  [arXiv:1509.02220 [hep-ph]].

\bibitem{Cirigliano:2008wn}
  V.~Cirigliano {\it et al.},
  JHEP {\bf 0811} (2008) 006
  [arXiv:0807.4507 [hep-ph]].

\bibitem{FlaviaNet}
  M.~Antonelli {\it et al.} [FlaviaNet Coll.],
  Eur.\ Phys.\ J.\ C {\bf 69} (2010) 399
  [arXiv:1005.2323 [hep-ph]].

\bibitem{Moulson:2014cra}
  M.~Moulson,
  arXiv:1411.5252 [hep-ex].

\bibitem{Bernard:2009zm}
  V.~Bernard {\it et al.},
  Phys.\ Rev.\ D {\bf 80} (2009) 034034
  [arXiv:0903.1654 [hep-ph]].

\bibitem{Callan:1966hu}
  C.~G.~Callan and S.~B.~Treiman,
  Phys.\ Rev.\ Lett.\  {\bf 16} (1966) 153.

\end{thebibliography}
\end{document}